\documentclass[a4paper, 12pt]{jpconf}
\usepackage{graphicx}
\usepackage{subfig}
\usepackage{cite}

\usepackage{enumerate}

\usepackage{float}

\usepackage{amsgen}
\usepackage{amsfonts}
\usepackage{amsbsy}
\usepackage{amssymb}

\usepackage{amsmath}
\usepackage{amsthm}

\usepackage{soul}
\setul{0.4pt}{.4pt}

\begin{document} 

\title[Effective lambda-nucleon potentials, inverse scattering ]{Effective lambda-proton and lambda-neutron potentials from subthreshold inverse scattering }
\author{E F Meoto and M L Lekala}
\address{Department of Physics, University of South Africa, \\ Private Bag X6, 1710 Johannesburg, South Africa}
\eads{\mailto{emilemeoto@aims.ac.za}, \mailto{lekalml@unisa.ac.za}}

\begin{abstract} 
Potentials are constructed for the lambda-nucleon interaction in the $^1\text{S}_0$ and $^3\text{S}_1$ channels. These potentials are recovered from scattering phases below the inelastic threshold through Gel'fand-Levitan-Marchenko theory. Experimental data with good statistics is not available for lambda-nucleon scattering. This leaves theoretical scattering phases as the only option through which the rigorous theory of quantum inverse scattering can be used in probing the lambda-nucleon force. Using rational-function interpolations on the theoretical scattering data, the kernels of the Gel'fand-Levitan-Marchenko integral equation become degenerate, resulting in a closed-form solution. The new potentials restored, which are shown to be unique through the Levinson theorem, bear the expected features of short-range repulsion and intermediate-range attraction. Charge symmetry breaking, which is perceptible in the scattering phases, is preserved in the new potentials. The lambda-nucleon force in the $^1\text{S}_0$ channel is observed to be stronger than in the $^3\text{S}_1$ channel, as expected. In addition, the potentials bear certain distinctive features whose effects on hypernuclear systems can be explored through Schr\"{o}dinger calculations. 
\end{abstract}

\section{Introduction}

The development of accurate potentials for hyperon-nucleon and hyperon-hyperon interactions is of prime importance in hypernuclear physics. Their importance lies in the extra quantum number that hyperons (e.g. lambda, sigma or cascade) introduce into nuclear systems. The nonzero strangeness quantum number of hyperons allows hypernuclei to occupy states that would be Pauli-forbidden if they contained only nucleons. Hence, these hyperons can be seen as probes for such \textit{genuine} hypernuclear states. Furthermore, some important phenomena within hypernuclei require accurate hyperon-nucleon and hyperon-hyperon potentials for a clearer understanding. Examples of such phenomena include the nonmesonic decay of hyperons in light hypernuclei and the reduction of nuclear size arising from the \textit{glue effect} of hyperons. Finally, these potentials are required in simulations of astrophysical objects with multistrangeness, for example the high-density core of a neutron star. 

In early helium bubble-chamber experiments and recent emulsion experiments, a large number of lambda hypernuclei have been observed, compared to just one or two sigma and cascade (xi) hypernuclei \cite{dav2005, dal2005}. As a result of this large number of lambda hypernuclei, the lambda-nucleon interaction has received considerable attention over the years. The lambda-nucleon potentials in common use are derived from meson-exchange theory \cite{deSwart1971, deSwart1996, rij1993,rij2001, hol1989,reu1992,hai2005} and quark theories \cite{fuj1996a, fuj1996b, fuj2001}. Some are based entirely on Quantum Chromodynamics, for example, Chiral Effective Field Theory \cite{pol2006, pol2007, hai2013}. The accuracy of these potentials have been tested by using them in few-body calculations to compute some important structural properties of light hypernuclei. These properties include the binding energy and lifetime of a lambda hypertriton. In some cases, charge symmetry breaking and lambda-sigma conversion, which are very important in the lambda-nucleon force, were also verified by computing the lambda separation energies of isospin multiplets. Differences, some negligibly small and others significant, were observed between these computations and experimental observations. For example, the computed lifetimes of the lambda hypertriton using existing lambda-nucleon potentials are about 30 - 50\% longer \cite{gal2018b} than the recently observed values \cite{ada2018}. This kind of significant differences point to the fact that much effort is still needed in understanding the lambda-nucleon force. All of this effort is currently invested in improving the application of particle-exchange theories and Quantum Chromodynamics theories on the lambda-nucleon force. In this paper, the approach is to probe the lambda-nucleon force through an existing theory, quantum inverse scattering, that has never been applied in this sector of the baryon-baryon force. The aim of the paper is to construct new lambda-neutron and lambda-proton potentials through Gel'fand-Levitan-Marchenko theory \cite{mar1950, mar1955, agr1963}. This theory is based on quantum inverse scattering at fixed angular momentum. A rigorous mathematical foundation for inverse scattering theory was established between the 1940s and 1960s. 

The rest of the paper has the following organisation: Sections 2 and 3 carry a brief outline of the problem statement in quantum inverse scattering and Gel'fand-Levitan-Marchenko theory, respectively. Section 4 presents an interpolation technique for the scattering matrix, which is important for the separability of the kernels appearing in the Gel'fand-Levitan-Marchenko integral equations. Sections 5, 6 and 7 discuss various aspects of the lambda-nucleon scattering data. Section 8 carries the results of the application of Gel'fand-Levitan-Marchenko theory to lambda-nucleon scattering. Concluding remarks are presented in section 9.   

\section{Quantum inverse scattering}

The recovery of a Sturm-Liouville operator from its spectral properties is a problem that arises in many contexts within the mathematical sciences. Developments on the theory of inverse Sturm-Liouville problems were pioneered in the 1940s by Borg \cite{bor1946,bor1949} and Povzner \cite{pov1948}. Applications of this theory to quantum scattering emerged later on through the work of Levitan \cite{lev1949}, Bargmann \cite{bar1949,bar1949b} and Levinson \cite{lev1949b,lev1949c}, among others. In quantum scattering, this problem arises in the restoration of the Schr\"{o}dinger operator from observed scattering data. For cases with spherical symmetry, the inverse scattering problem is an inverse Sturm-Liouville problem for the Schr\"{o}dinger operator on the half-line \cite{lev1987}:

\begin{align} 
L^{(\ell)} \psi = E \psi, \quad 0 \leq r < \infty
\end{align} 
where
\begin{align} 
L^{(\ell)} = - \frac{d^2}{dr^2} + V(r)+\frac{\ell(\ell+1)}{r^2}
\end{align} 

$\psi$ is the radial wavefunction, $\ell$ is the orbital angular momentum number and $V(r)$ is the potential. The system of units used is such that $\hbar = 2 \mu = 1$, where $\mu$ is the reduced mass of the system. In these units, the energy is given by $E=k^2$ and the momentum by $\vec{p}=\vec{k}$, where $\vec{k}$ is the wavevector. The aim of this study is to recover the operator $L^{(\ell)}$ from available scattering phases, $\delta_{\ell}(k)$. This is carried out for $V(r)$ being lambda-proton and lambda-neutron potentials. 

\section{Gel'fand-Levitan-Marchenko theory}

Transformation operators for solutions of the Schr\"{o}dinger equation play a central role in inverse scattering theory. Originally developed in shift operator theory by Delsarte \cite{del1938a, del1938b}, transformation operators were introduced in inverse scattering theory by Marchenko \cite{mar1950, mar1952}. The application of these transformation operators on the solutions of the Schr\"{o}dinger equation results in a Povzner-Levitan integral representation for these solutions \cite{pov1948,lev1949}. From the Povzner-Levitan representation for the Jost solutions, one arrives at the single-channel Gel'fand-Levitan-Marchenko (GLM) equation \cite{mar1955,agr1963}:
\begin{align}
\label{eq:marchenko}
 K_{\ell}(r,r') + A_{\ell}(r,r') + \int_r^\infty  K_{\ell}(r,s) A_{\ell}(s,r') ds = 0, \quad r < r'
\end{align}
where $K_{\ell}(r,r')$ and $A_{\ell}(r,r')$ are the kernels of the integral equation. $K_{\ell}(r,r')$ is related to the potential $V(r)$ through a hyperbolic differential equation \cite{coz1973, coz1987}. $A_{\ell}(r,r')$ is computed from the continuum and discrete spectra as follows \cite{mas1999, ben1969,ben1972,new2002, cha1977}:
\begin{align}
\label{eq:inputkernel2}
 A_{\ell}(r,r')= \frac{1}{2 \pi} \int_{-\infty}^{\infty} \omega_{\ell}^{+}(kr) \left\{ 1 - S_{\ell}(k)   \right \} \omega_{\ell}^{+}(kr') dk + \sum_{i=1}^{n_{\ell}} M_{i \ell} \omega^{+}_{\ell}(k_i r) \omega^{+}_{\ell}( k_i r')
\end{align}
$S_{\ell}(k)$ is the partial-wave scattering matrix, $\omega_{\ell}^{+}(kr)$ are outgoing Riccati-Hankel functions, $n_{\ell}$ is the number of bound states, $k_i$ are bound state wavenumbers and $M_{i \ell}$ are norming constants for the bound states. The first term in Equation \eqref{eq:inputkernel2} is the contribution from the positive eigenvalues (continuum spectrum) of the Schr\"{o}dinger operator while the second term is the contribution from the negative eigenvalues (discrete spectrum). The ingoing and outgoing Riccati-Hankel functions are defined as follows \cite{cha1977}:
\begin{align}
\omega_{\ell}^{\pm}(kr) = \sqrt{\frac{\pi k r}{2}} H^{\pm}_{\ell+1/2}(kr)
\end{align}
where $H^{\pm}_{\ell+1/2}(kr)$ are Hankel functions. Generally, there is no guiding physical law for computing fixed values of the norming constants $M_{i \ell}$. Therefore, in cases where there are negative eigenvalues, instead of a unique potential, one ends up with a set of phase-equivalent potentials \cite{sof1990}, which may be isospectral \cite{lev1949c}. As outlined in \cite{mas1999, new2002}, one method used to still get a unique solution when there are bound states is by computing fixed values of the constants $M_{i \ell}$ from the Jost solutions. 
 
After computing $A_{\ell}(r,r')$ from the input data, the GLM equation is solved to obtain the output kernel $K_{\ell}(r,r')$. From the boundary conditions for the Goursat problem satisfied by $K_{\ell}(r,r')$ \cite{coz1973, lev1987}, the potential $V_{\ell}(r)$ is obtained through the diagonal entries in $K_{\ell}(r,r')$, i.e.
\begin{align}
\label{ch:potentialsolution}
-2\frac{d }{d r}K_{\ell}(r,r) &= V_{\ell}(r)
\end{align}
The separability of the kernels $A_{\ell}(r,r')$ and $K_{\ell}(r,r')$ determine whether the GLM integral equation will have a closed-form solution or a numerical algorithm is needed. In the following section, the properties of the scattering matrix and how they affect the separability of these kernels are discussed. From this point forward, the discussion shall be restricted to the case $\ell=0$ (the $s$-waves), which is of interest in this paper.

\section{Separability of the kernels} 

If the kernels of an integral equation are separable, an exact solution can generally be found this equation.  In order to achieve separability in the kernels of the GLM integral equation, the scattering matrix is interpolated by a suitable function $\tilde{S}_{0}(k)$. From Bargmann rational-function representations of the Jost function \cite{bar1949}, a suitable approximation for the scattering matrix is as follows \cite{har1981, mas1999,rak2007}: 
\begin{align}
\tilde{S}_{0}(k)= \prod_{n=1}^{N} \left (\frac{k + \alpha^{0}_n}{k - \alpha^{0}_n} \right ) \left ( \frac{k - \beta^{0}_n}{k + \beta^{0}_n} \right )
\end{align}
where $\alpha^{0}_n$ and $\beta^{0}_n$ are complex numbers representing the zeros and poles of the Jost functions. For uniformity, one may use the same symbol, $a^{0}_m$, to represent all the $\alpha^{0}_n$'s and $\beta^{0}_n$'s. In cases where there are no bound states, the approximation therefore takes the following simple form \cite{har1981, mas1999,rak2007}:
\begin{align}
\label{eq:smatrix_final}
\tilde{S}_{0}(k) = \prod_{m=1}^{M} \frac{k + a^{0}_m}{k - a^{0}_m}
\end{align}
where $M$ is an even number. In the available scattering data, the momenta, $k$, are real. The rational function representation of the scattering matrix in Equation \eqref{eq:smatrix_final} ensures an analytic continuation of the domain to complex momenta i.e. $k= x + i y$, where $x$ and $y$ are real, with $i^2=-1$. The complex constants $a^{0}_m$ are estimated by rewriting the rational function in Equation \eqref{eq:smatrix_final} as a Pad\'{e}  approximation of order $[M/M]$ \cite{har1981, mas1999,rak2007}. By selecting $M$ data points $(k_j, S_{0}(k_j))$ from the original scattering matrix data $S_{0}(k)$ and substituting into the Pad\'{e} approximation, a linear system is obtained, as shown in \cite{pap1990}. In the Pad\'{e}  approximation of order $[P/Q]$, the requirement $P=Q$ ensures that $\tilde{S}_{0}(k) \to 1$ as $k \to \infty$, a condition that is satisfied by short-range potentials. 

The rational-function interpolation for the scattering matrix causes the kernel $A_{\ell}(r,r')$ to become separable. Based on this degeneracy of $A_{\ell}(r,r')$, a separability ansatz is assumed for $K_{\ell}(r,r')$. With these separable kernels, there is no need for quadrature as an exact solution can be found for the GLM integral equation. This solution is outlined in \cite{pap1990, kub1987, kir1989}. A solution to the inverse scattering problem with degenerate kernels was earlier derived by Faddeev \cite{fad1959,fad1963}. Another advantage of $\tilde{S}_{0}(k)$ is that it can be used over a wider momentum range than that covered by the original scattering data, $S_{0}(k)$. Extrapolation to higher momenta is necessary in some cases to ensure that the problem is well-posed. In the following sections, single-channel GLM theory, as outlined up to this point, is applied to lambda-nucleon scattering. 

\section{Theoretical lambda-nucleon scattering data below threshold}
\label{sec:phases}

In the kinematics of lambda-nucleon scattering experiments, free $\Lambda$ hyperons are usually used as projectiles. These experiments are very difficult to perform due to the very short lifetime of hyperons, which is about $2.63 \times 10^{-10}$ seconds \cite{oli2014}. The reverse kinematics is also expected to be difficult for the same reason. Of interest in this paper is elastic scattering, for which the number of particles is conserved. In elastic scattering, both the $\Lambda$ and the nucleon emerge in final quantum states that are the same as their initial states:
\begin{align}
\Lambda + p &\to \Lambda + p \\
\Lambda + n &\to \Lambda + n 
\end{align}
Whereas the nucleon-nucleon scattering database has about 4000 data points, the lambda-nucleon database has only about 40 data points. Experiments on hyperon-hyperon scattering, which are even more difficult, have never been reported. In addition to the low number of data points, some of the hyperon-nucleon data sets have large error bars. Furthermore, the number of lambda-nucleon scattering events is too low for any decent application of inverse scattering theory. As a result of the limited experimental scattering data, one therefore has to resort to using theoretical or simulated data, so that the powerful theory of inverse scattering can be employed in probing the lambda-nucleon force. The use of theoretical scattering data is known to have contributed in elucidating the nature of the nucleon-nucleon interaction. For example, in \cite{kir1989} phase shifts computed from the Reid soft core potential were used in restoring the nucleon-nucleon potential. 

\begin{figure}[!h]
\centering
\subfloat[$\Lambda p$ phase shifts in the $^1\text{S}_0$ and $^3\text{S}_1$ channels.]{\includegraphics[width=0.7\linewidth]{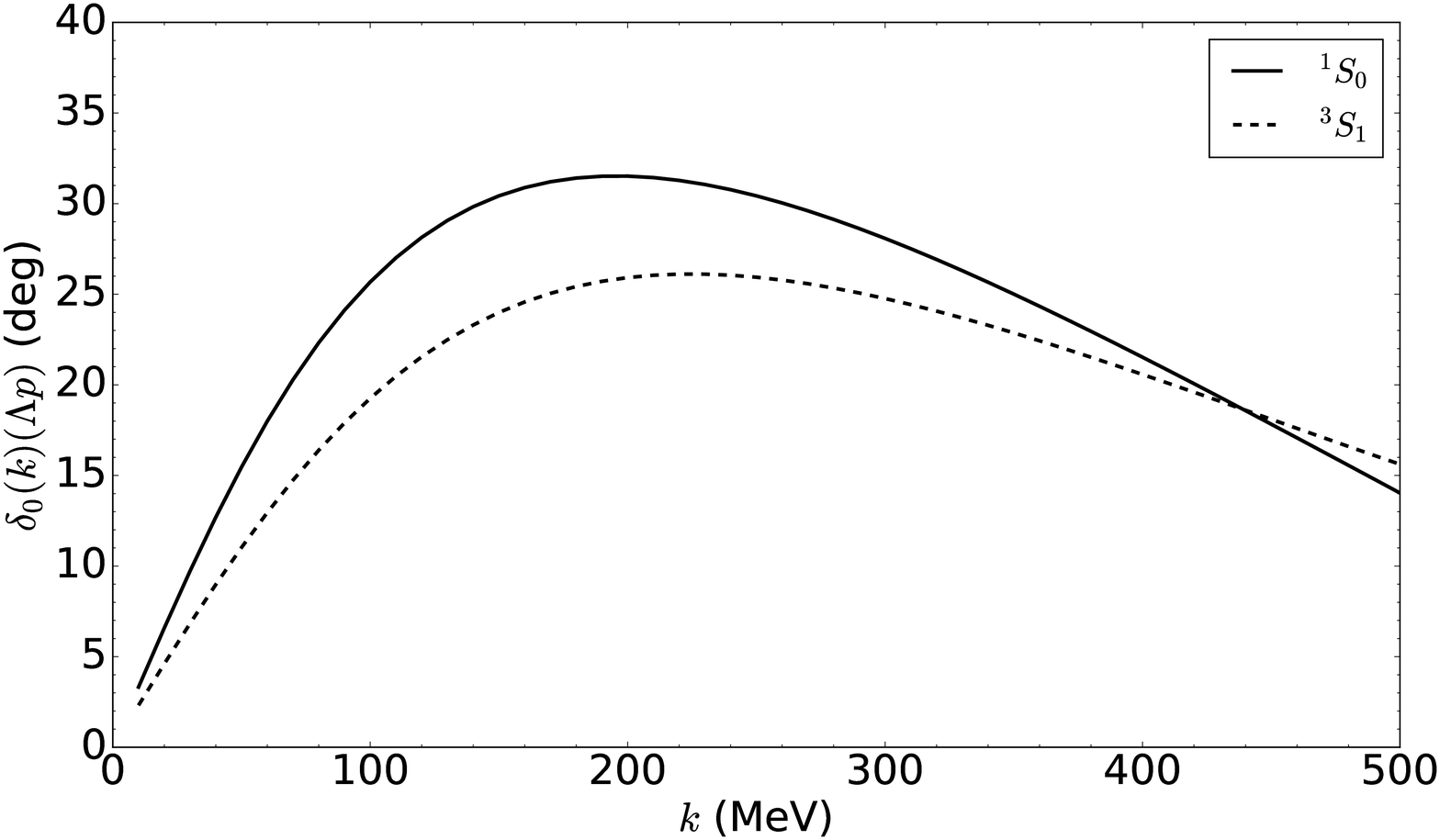}\label{fig:phase_shift_lp_both}}

\subfloat[$\Lambda n$ phase shifts in the $^1\text{S}_0$ and $^3\text{S}_1$ channels.]{\includegraphics[width=0.7\linewidth]{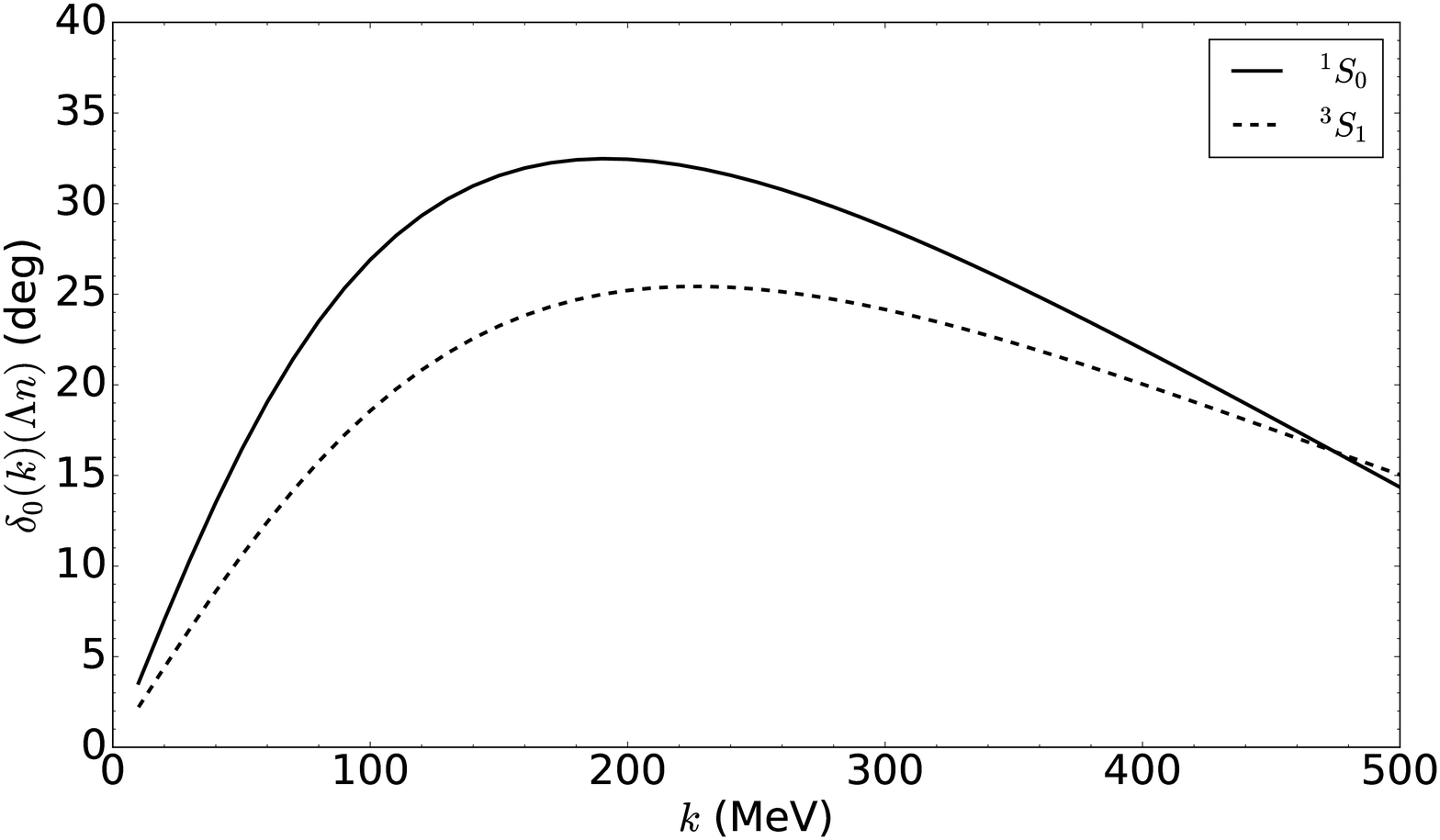}\label{fig:phase_shift_ln_both}}
\caption{Theoretical lambda-nucleon phase shifts below the inelastic threshold. The difference between the $\Lambda p$ and $\Lambda n$ scattering phases is barely noticeable because of the very small mass difference between a proton and neutron.}
\label{fig:phase_shift}
\end{figure}

In this paper, theoretical $^1\text{S}_0$ and $^3\text{S}_1$ phase shifts computed by the Nijmegen group \cite{ren2017} are used in our application of GLM theory. These theoretical $\Lambda p$ and $\Lambda n$ phase shifts, which were computed using the NSC97f potential \cite{rij1999}, are shown in Figures \subref{fig:phase_shift_lp_both} and \subref{fig:phase_shift_ln_both}, respectively. From this theoretical data, the inelastic threshold for $\Lambda p$ scattering is observed to be 640 MeV while that for $\Lambda n$ scattering is 650 MeV. This $\Lambda p$ threshold is almost the same as that observed in experiments, for example in \cite{hau1977}. Since no $\Lambda n$ experimental scattering data has been reported, it is not possible to compare the theoretical threshold. The scattering phases used in our application cover the momentum range up 500 MeV, which is below the inelastic threshold. Due to the fact that there is no loss of flux arising from inelastic channels, these subthreshold phase shifts have no imaginary part.

\section{Levinson theorem}

In preparation for solving the GLM equation, the Levinson theorem was used to examine the scattering phases in Figure \ref{fig:phase_shift}, in order to determine if there are any bound states. For scattering by a short-ranged potential with spherical symmetry, when $\ell=0$, the Levinson theorem is given by the following relation \cite{cha1977}:
\begin{align}
\delta_{0}(0)-\delta_{0}(\infty) =  \pi n_{0}
\end{align} 
where the constant $n_{0}$ is the number of bound states. 

In Figure \ref{fig:phase_shift}, it can be observed that $\delta_{0}(k) \to 0$ as $k \to 0$ for all the phase shifts. The behaviour of the phase shift at infinity can be inferred from the nature of the potentials. The scattering theory applied in this paper is valid for short-range potentials. The scattering matrix must therefore behave in such a way that $S_{0}(k) \to 1$ as $k \to \infty$. Or, equivalently, the phase shift must vanish at very high momenta i.e. $\delta_{0}(k) \to 0$ as $k \to \infty$. Therefore, based on the application of the Levinson theorem on these phase shifts, the lambda-nucleon potentials do not support any bound states. 

\section{Distribution of the poles of the scattering matrix approximation}

The scattering phases are related to the partial-wave scattering matrix through the relation $S_{\ell}(k)= \eta_{\ell}\text{exp}(2i\delta_\ell(k))$. For a single-channel problem, $\eta_{\ell} = 1$ below the inelastic threshold, and the scattering matrix is unitary. 

\begin{figure}[h]
\centering
\subfloat[]{\includegraphics[width=0.6\linewidth]{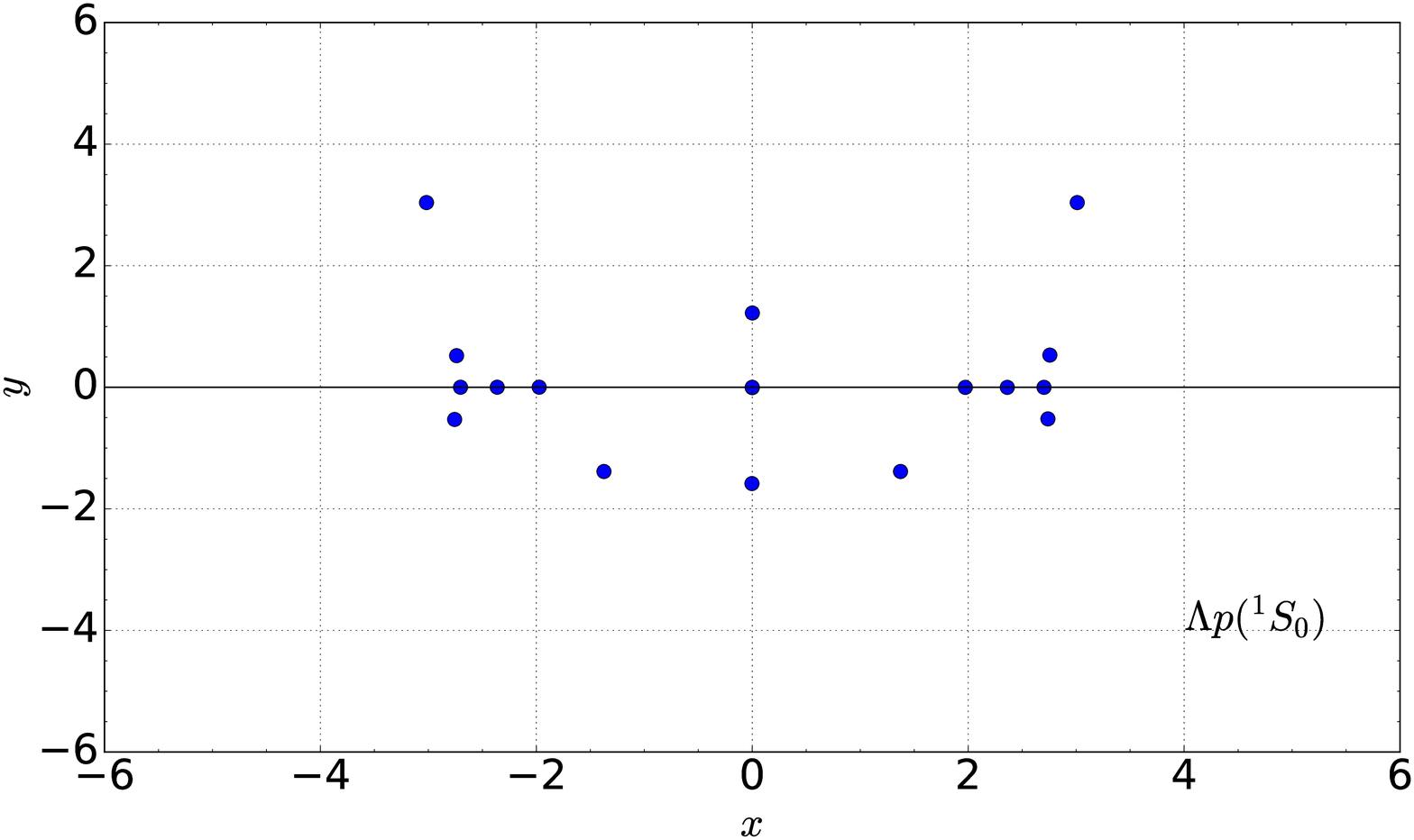}\label{fig:poles_lp1s0}}

\subfloat[]{\includegraphics[width=0.6\linewidth]{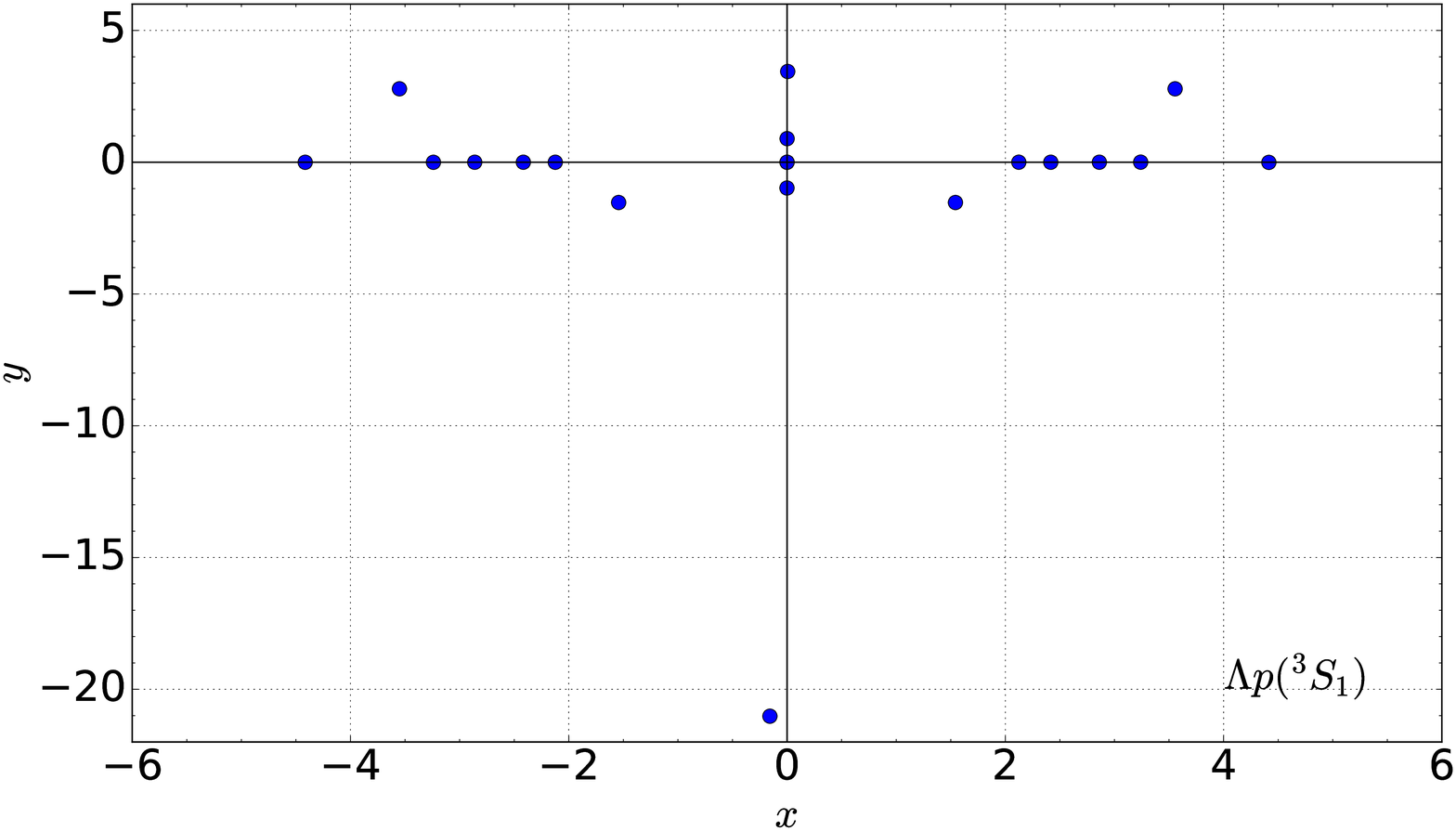}\label{fig:poles_lp3s1}}
\caption{Distribution of poles in $\Lambda p$ scattering matrix approximation.}
\label{fig:poles_lp}
\end{figure} 

\begin{figure}[!h]
\centering
\subfloat[]{\includegraphics[width=0.6\linewidth]{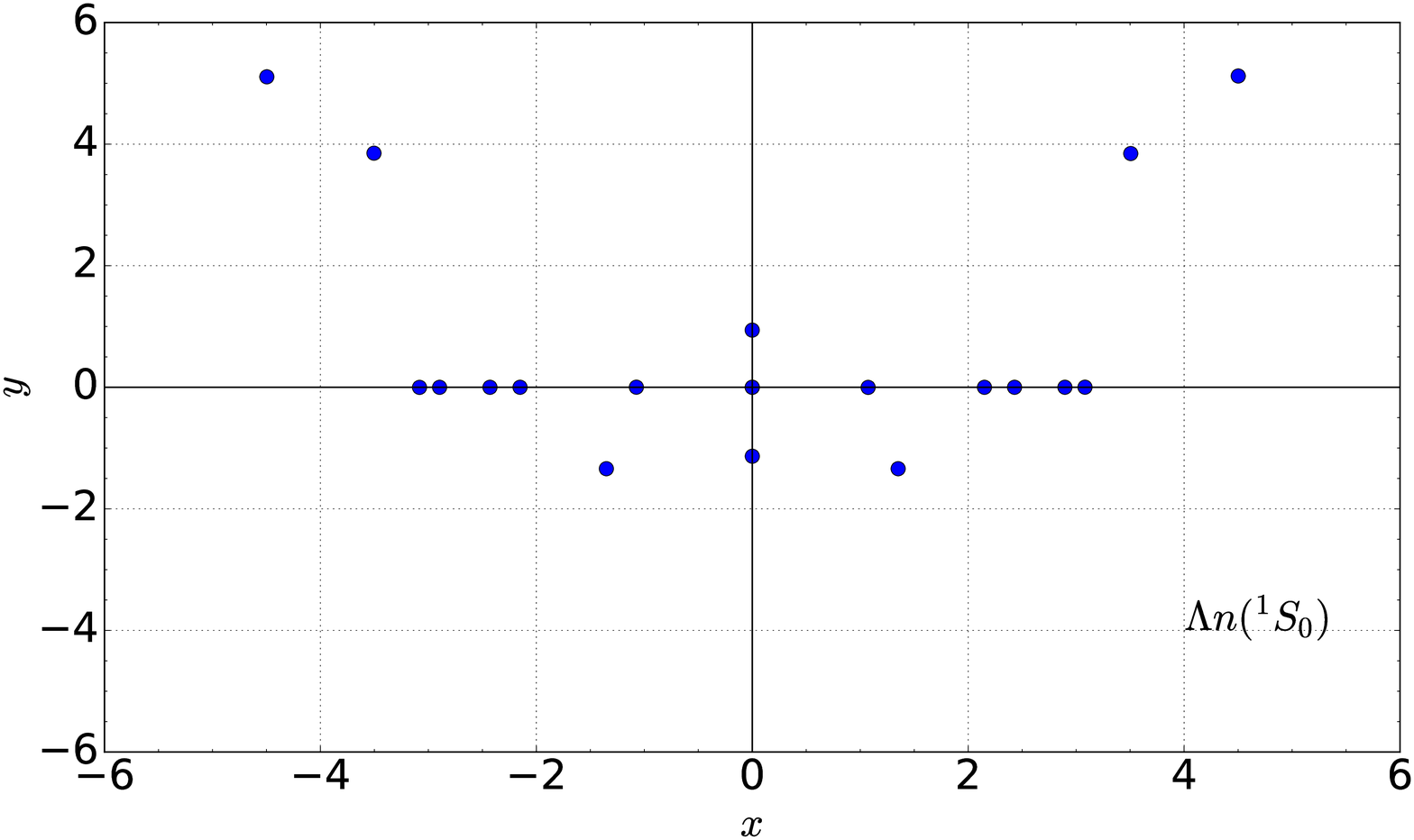}\label{fig:poles_ln1s0}}

\subfloat[]{\includegraphics[width=0.6\linewidth]{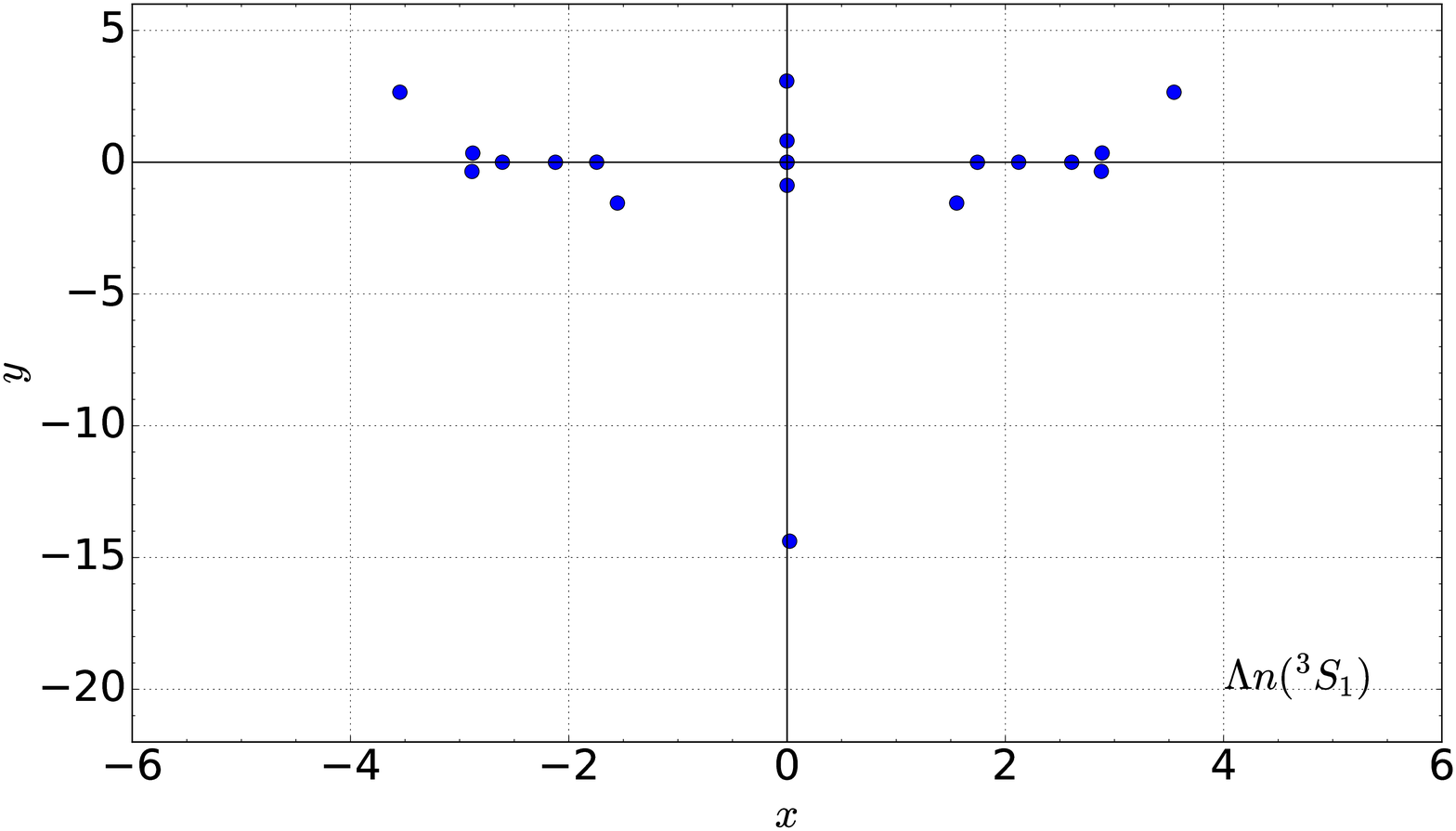}\label{fig:poles_ln3s1}}
\caption{Distribution of poles in $\Lambda n$ scattering matrix approximation.}
\label{fig:poles_ln} 
\end{figure} 

The scattering phases in Section \ref{sec:phases} were used to compute scattering matrices through the relation $S_{0}(k)=\text{exp}(2i\delta_{0}(k))$. Next, the approximation in Equation \eqref{eq:smatrix_final} was used to compute $\tilde{S}_{0}(k)$. In the numerical method used in estimating the $M$ constants $a^{0}_m $, as described in \cite{mas1999,rak2007}, values of $M \in [4,20]$ are known to ensure sufficient accuracy for any practical applications \cite{kir1989}. For example, good approximations for nucleon-nucleon and nucleon-deuteron scattering data were obtained in \cite{coz1987, rak2007, pap1990} using $M=6, 10, 20$ poles, respectively. In \cite{kir1989}, a good approximation was obtained with $M=8$ or $M=12$ per 23 data points. The scattering matrix is directly related to the effective range function, $R_{\ell}(k^2)=k^{2 \ell+1} \cot \delta_{\ell}(k)$, whose series expansion has a finite radius of convergence. Since this radius of convergence is usually small, $M$ cannot be made arbitrarily large.

For the approximation carried out in this paper, twenty poles ($M=20$) were used for $\tilde{S}_{0}(k)$. This number of poles has been shown to accurately constrain scattering phases with a momentum dependence similar to those in this paper, for example in \cite{pap1990, alt1994}. The distribution of the computed poles of $\tilde{S}_{0}(k)$ in the complex momentum plane are shown in Figures \ref{fig:poles_lp} and \ref{fig:poles_ln}. It can be observed that the poles either lie on the imaginary axis ($x=0$) or have a symmetrical distribution about the imaginary axis. This is a confirmation of the unitarity of $\tilde{S}_{0}(k)$, a known property of $S_{0}(k)$ for subthreshold scattering \cite{kub1987}. Unitarity of the scattering matrix is a required condition for quantum inverse theory below the inelastic threshold. 

After analytic continuation to complex momenta, the poles of the scattering matrix hold information on the spectral points of the system. However, since Equation \eqref{eq:smatrix_final} is not a \textit{true} factorisation of the scattering matrix into the Jost functions and the number $M$ can vary, not all the poles of $\tilde{S}_{0}(k)$ are spectral points. Some of them are true poles of the original scattering matrix (or can be made to converge to the true poles by varying $M$) while others are spurious poles \cite{rak2007}. Therefore, before the spectral points can be investigated, the true poles must be separated from the spurious poles. This is a subject that can be attended to as a separate project. For purposes of using these poles in solving the GLM equation, the difference between true poles and spurious poles is insignificant.

\section{Results and discussion}

\subsection{Spin-dependent potentials}
\begin{figure}[!h]
\centering
\subfloat[]{\includegraphics[width=0.5\linewidth]{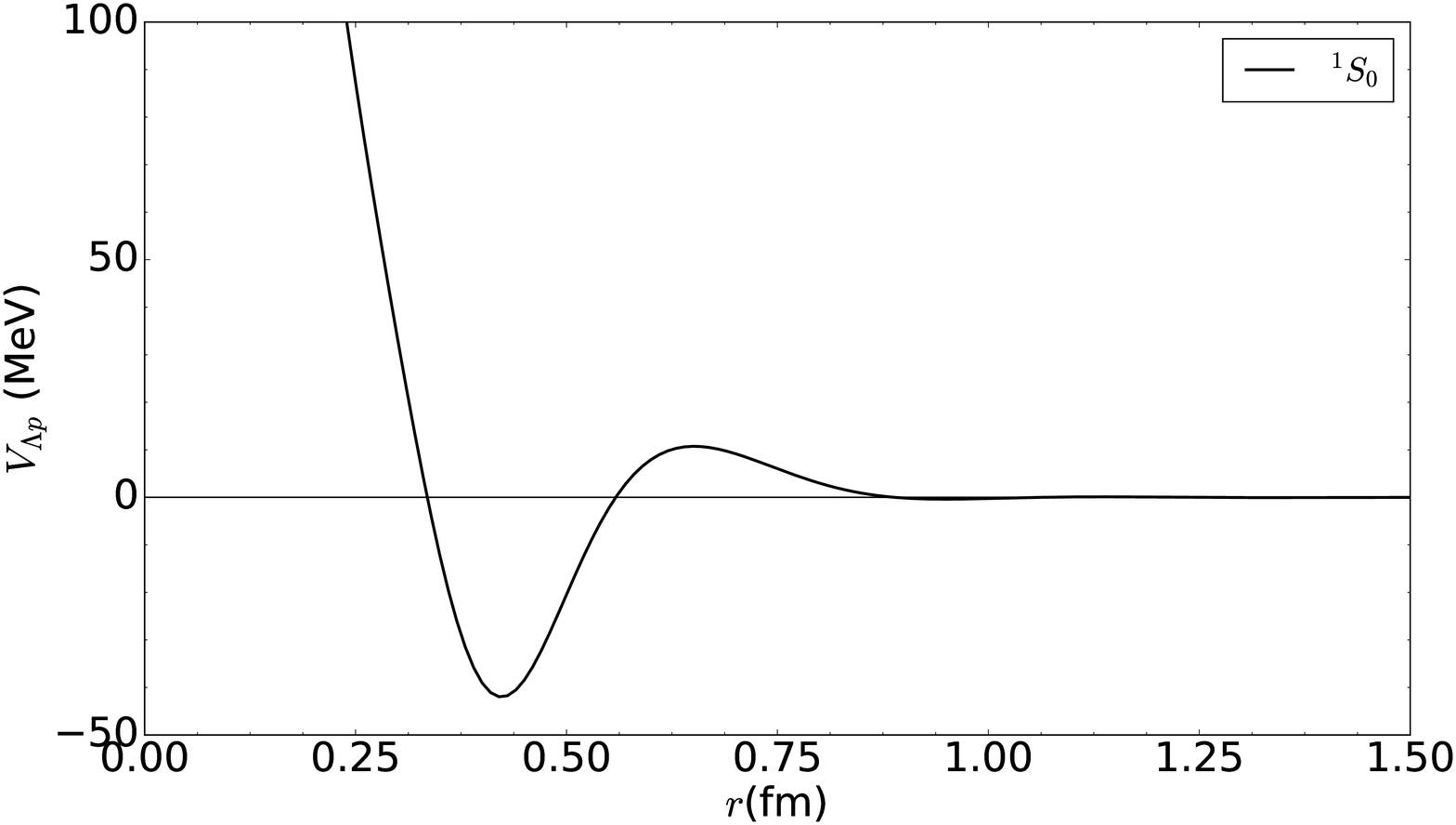}\label{fig:v_lp1s0}}
\subfloat[]{\includegraphics[width=0.5\linewidth]{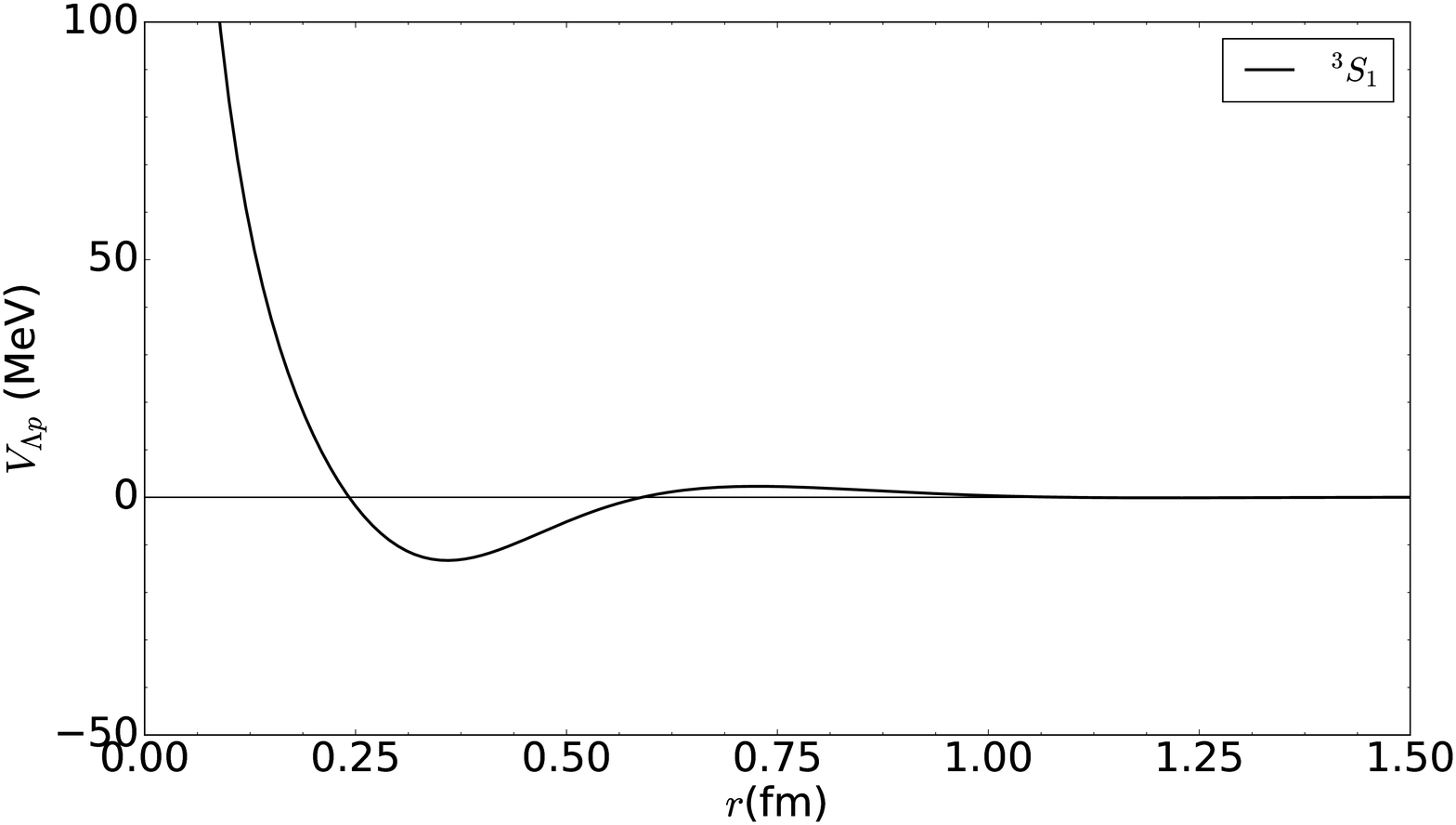}\label{fig:v_lp3s1}}
\caption{$\Lambda p$ potentials from GLM theory. The $^1\text{S}_0$ and $^3\text{S}_1$ attraction depths are -41.96 MeV and -13.28 MeV, respectively.}
\label{fig:v_lp}
\end{figure}

\begin{figure}[!h]
\centering
\subfloat[]{\includegraphics[width=0.5\linewidth]{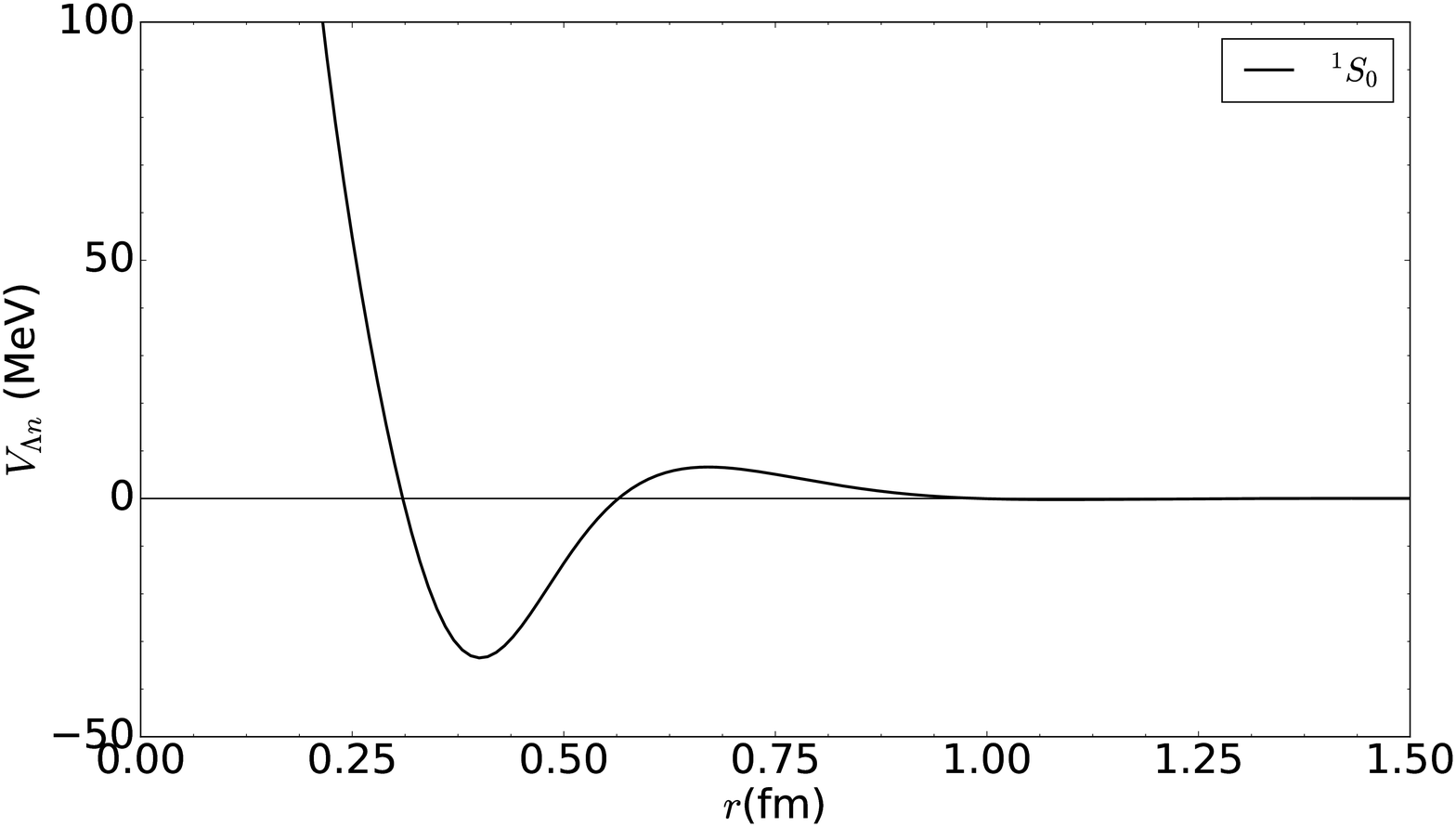}\label{fig:v_ln1s0}}
\subfloat[]{\includegraphics[width=0.5\linewidth]{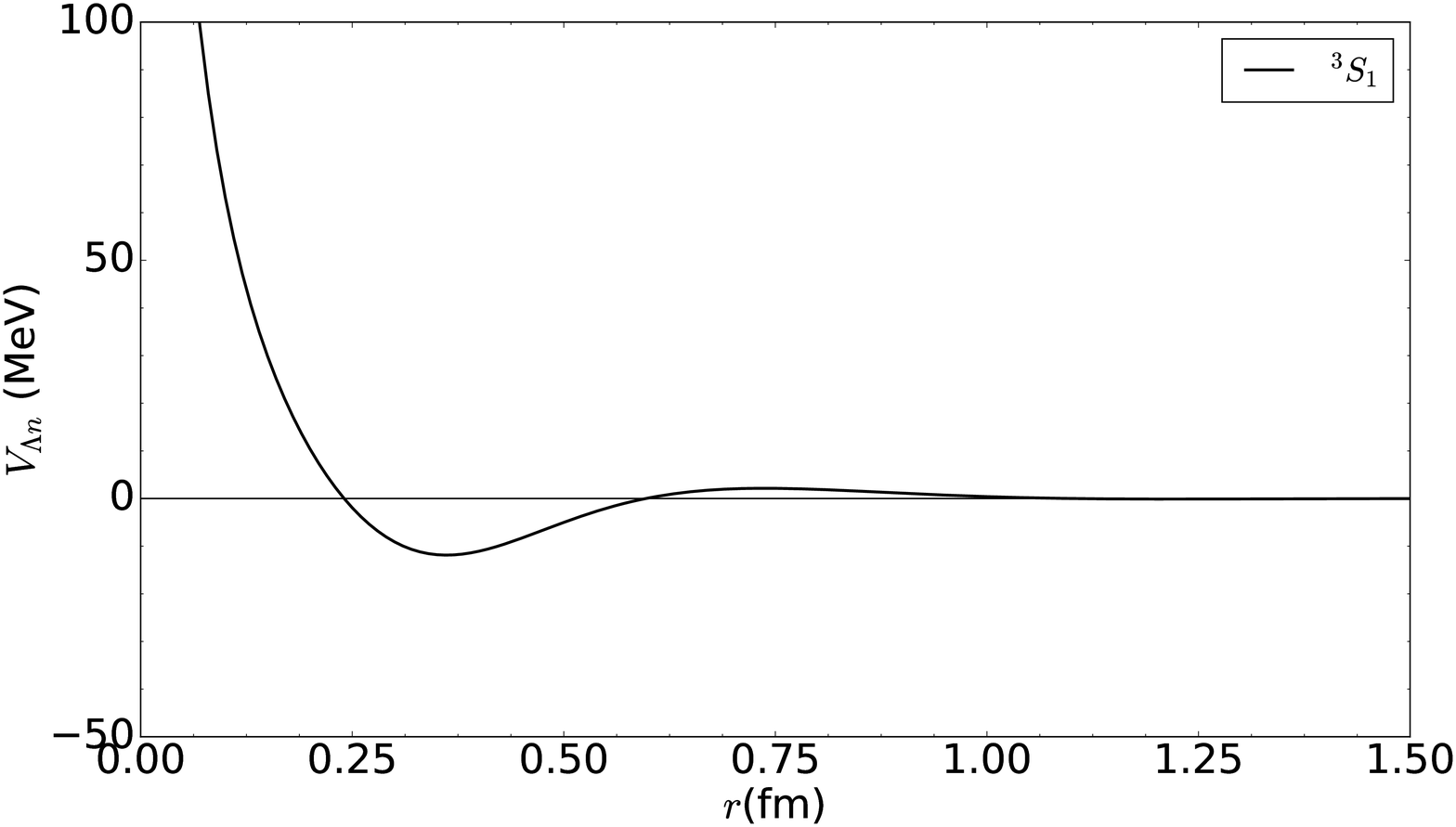}\label{fig:v_ln3s1}}
\caption{$\Lambda n$ potentials from GLM theory. The $^1\text{S}_0$ and $^3\text{S}_1$ attraction depths are -33.45 MeV and -11.87 MeV, respectively.}
\label{fig:v_ln} 
\end{figure}

Single-channel GLM theory was applied to the scattering phases from Section 5. Using $\tilde{S}_{0}(k)$, the integral in Equation \eqref{eq:inputkernel2} for the input kernel was computed through the Cauchy Residue Theorem as shown in \cite{amo1989,kir1989}. The GLM equation was solved as outlined in \cite{amo1989, kir1989,pap1990,alt1994}. The potentials obtained for the $\Lambda p$ interaction are shown in Figures \subref{fig:v_lp1s0} and \subref{fig:v_lp3s1}, while those for the $\Lambda n$ interaction are shown in Figures \subref{fig:v_ln1s0} and \subref{fig:v_ln3s1}. 

\begin{figure}[!h]
\centering
\subfloat[]{\includegraphics[width=0.5\linewidth]{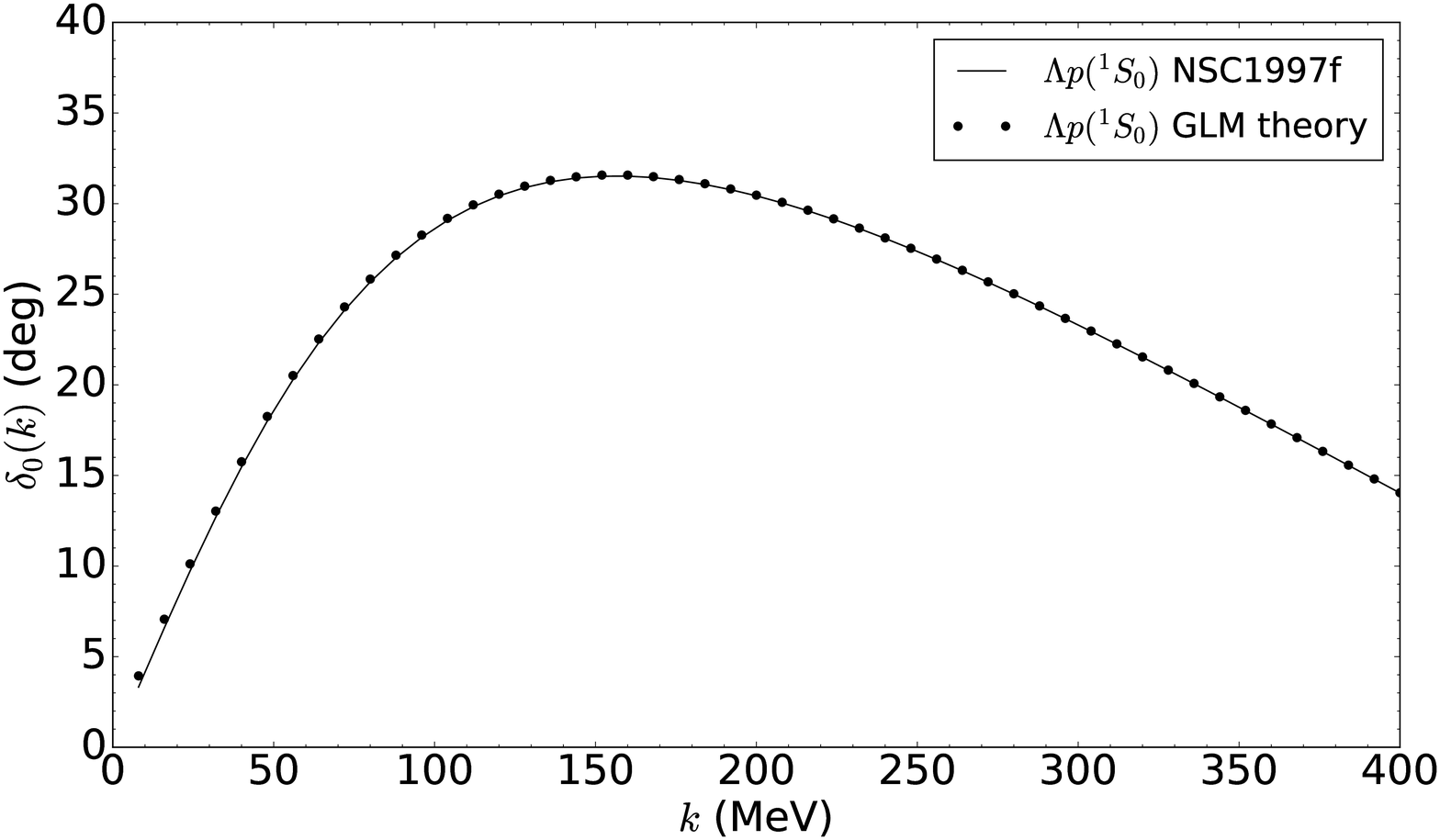}\label{fig:compare_lp1s0}}
\subfloat[]{\includegraphics[width=0.5\linewidth]{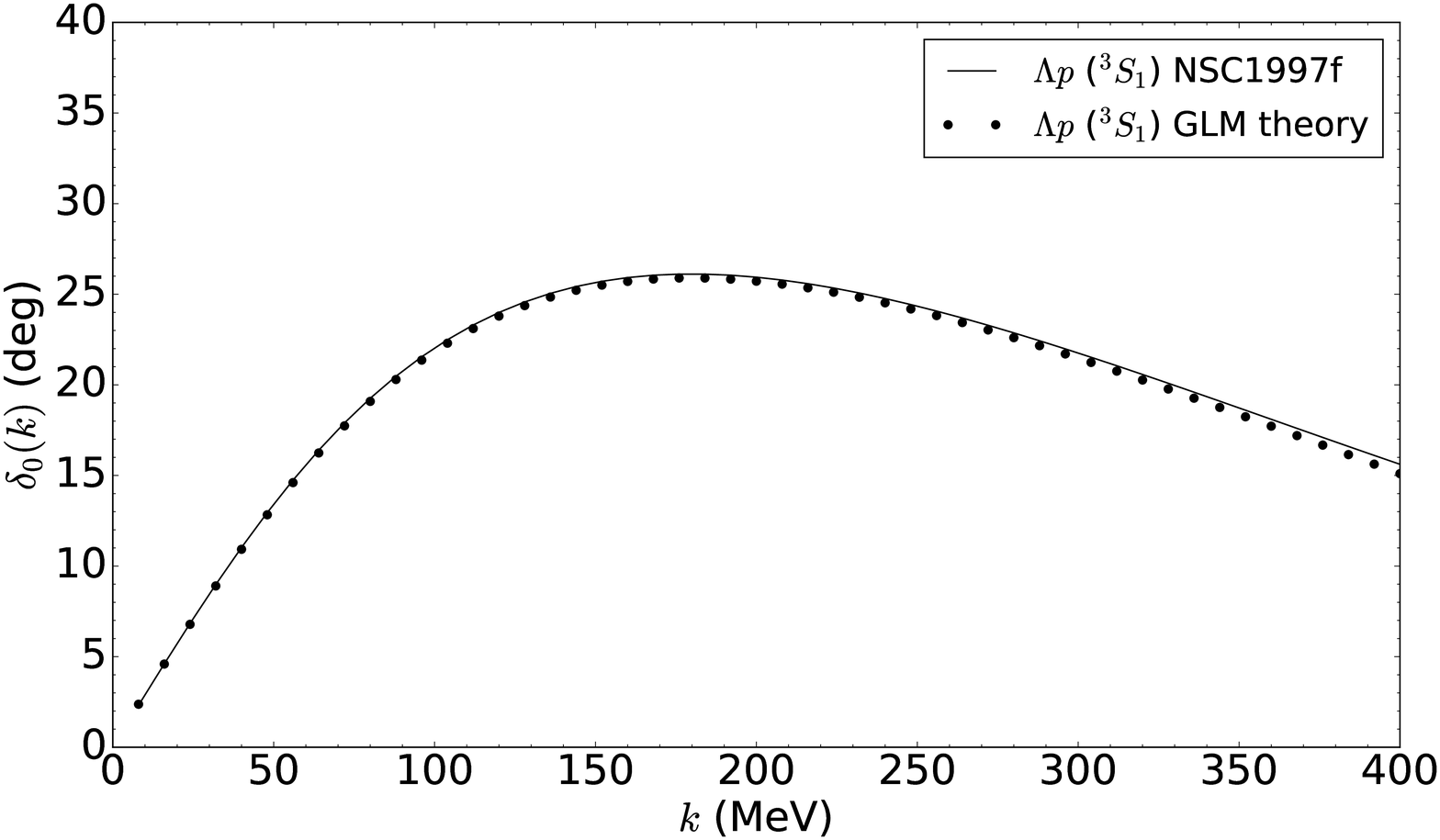}\label{fig:compare_lp3s1}}
\caption{$\Lambda p$ phase equivalence between NSC1997f and GLM theory potentials.}
\label{fig:compare_lp}
\end{figure} 

\begin{figure}[!h]
\centering
\subfloat[]{\includegraphics[width=0.5\linewidth]{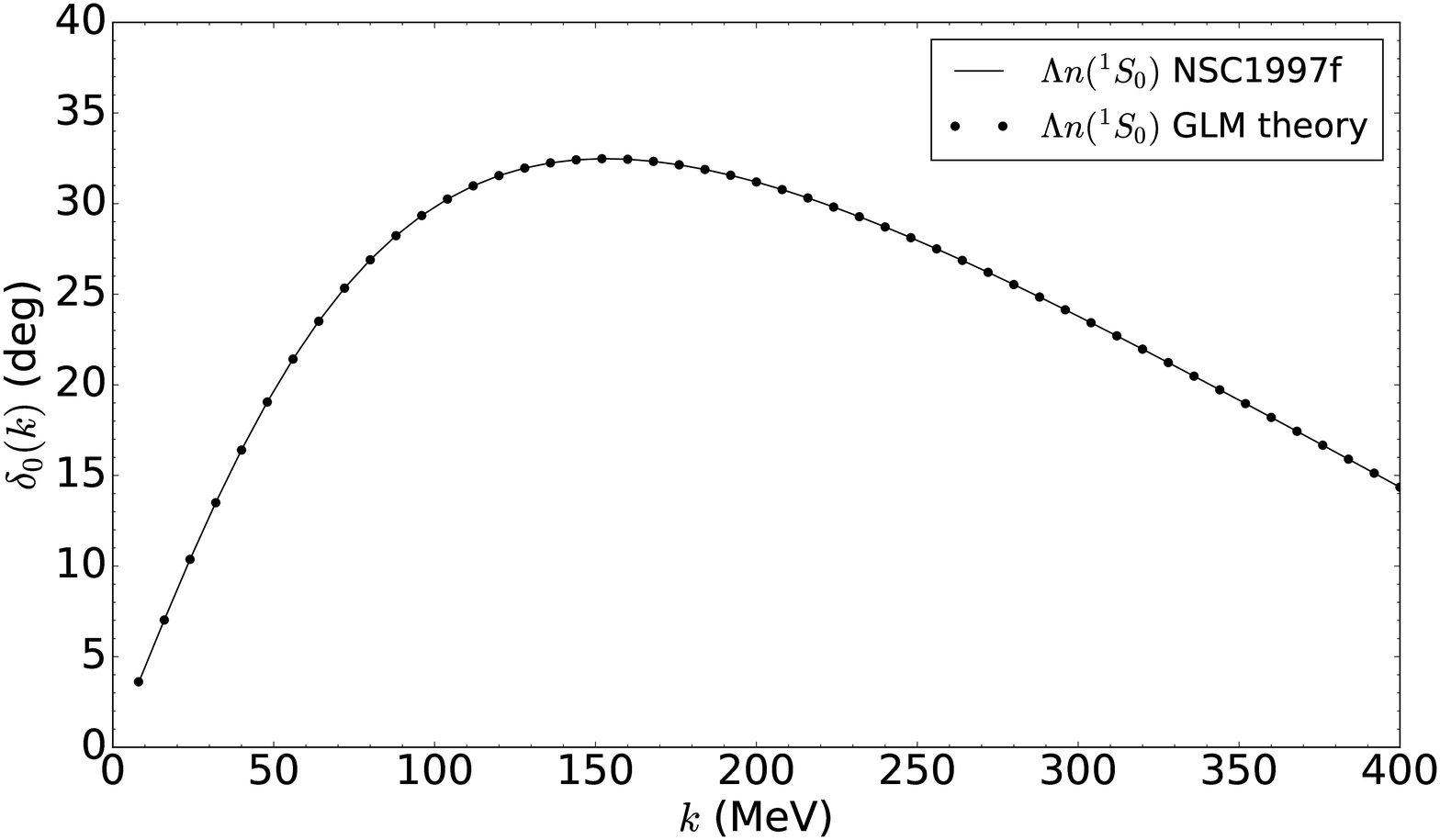}\label{fig:compare_ln1s0}}
\subfloat[]{\includegraphics[width=0.5\linewidth]{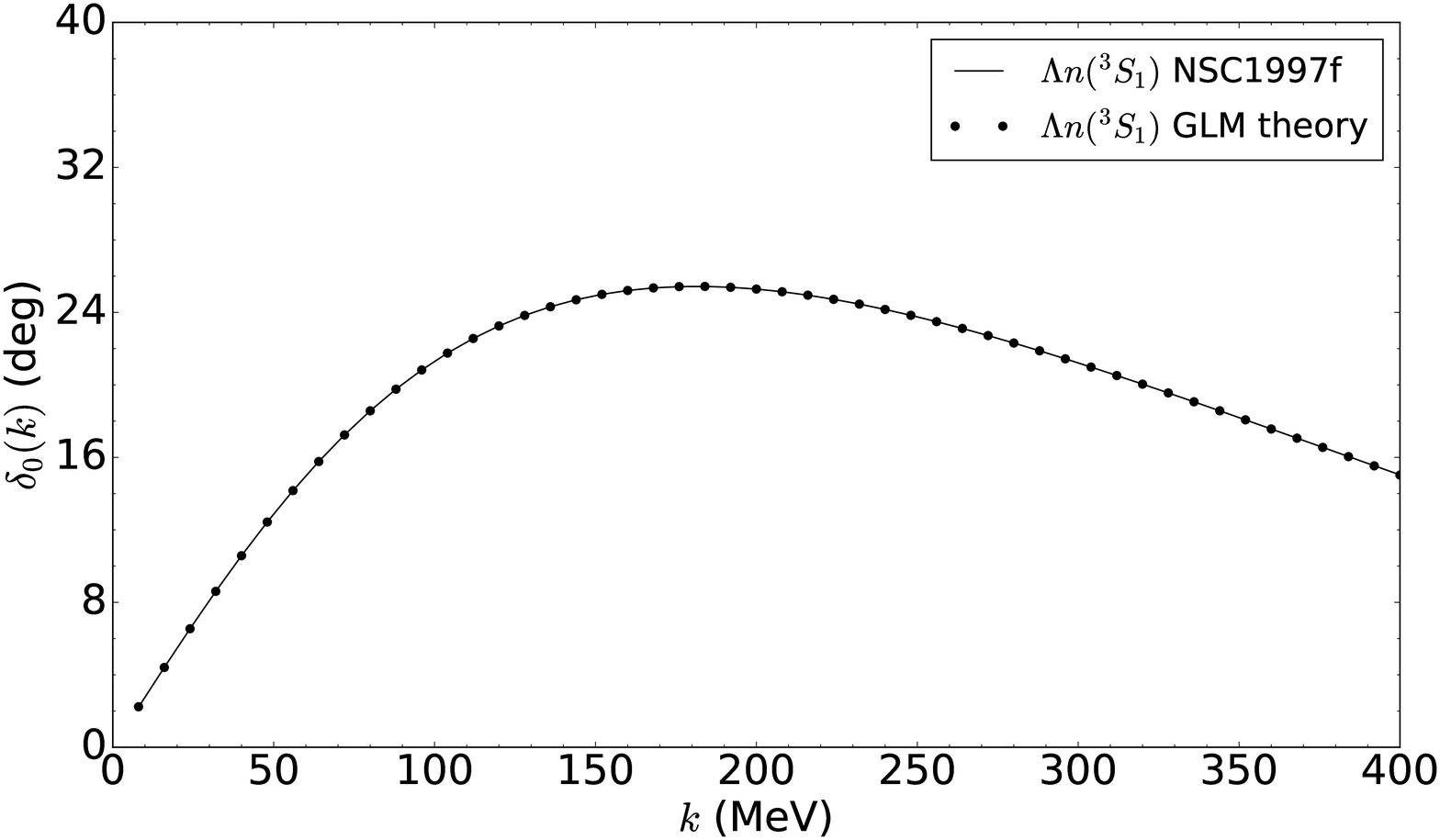}\label{fig:compare_ln3s1}}
\caption{$\Lambda n$ phase equivalence between NSC1997f and GLM theory potentials.}
\label{fig:compare_ln} 
\end{figure}

Observation of the results reveals that these new potentials bear the features of short-range repulsion and intermediate-range attraction, as expected for a baryon-baryon interaction. However, the attraction depth is located at a smaller radial distance when compared with other lambda-nucleon potentials, for example the simulation of the NSC97f potential in \cite{hiy2002}. This is a feature whose effects can be investigated through few-body calculations.

Just beyond the intermediate-range attraction a small repulsion barrier is noticeable, in the region $0.55 - 1.00$ fm. The barrier is observed to be consistently stronger in the $^1\text{S}_0$ channel than in the $^3\text{S}_1$ channel. Within inverse scattering theory, oscillations which are \textit{similar} to this repulsion barrier are often attributed to the effects of uncertainties or the onset of inelasticity in the input scattering data when the approximated scattering matrix is extrapolated to higher momenta \cite{kub1987,how1993}. An investigation on how uncertainties propagate into the potential from inverse scattering theory is found in \cite{kub1987, lee1992, adam1993b}. It is relevant to point out that such a repulsion barrier has also been observed in recent studies on some hyperon-nucleon potentials in meson theory \cite{nag2019}.

The restored potentials are spin-dependent, with the $\Lambda N(^1\text{S}_0)$ potential having a stronger attraction than the $\Lambda N(^3\text{S}_1)$ potential. This is a feature which is already known for the lambda-nucleon force. An explanation for this difference in strength between the $^1\text{S}_0$ and $^3\text{S}_1$ channels can only be provided by examining the strong force between the quark constituents of hyperons and nucleons. The short-range repulsion in the $\Lambda N(^3\text{S}_1)$ channels of these potentials is a correction on our earlier application of GLM theory to the lambda-nucleon force, presented in \cite{meo2017}. 

The $\Lambda p$ potential is slightly more attractive than the $\Lambda n$ potential, in both the $^1\text{S}_0$ and $^3\text{S}_1$ channels. This is known to arise from charge symmetry breaking in the lambda-nucleon force. The proton is constituted by two up quarks and one down quark (uud), the neutron by one up quark and two down quarks (udd). Due to the fact that the down quark is heavier than the up quark, coupled with the dynamics of quark-quark interactions inside hadrons, the neutron has a slightly higher mass than the proton. This mass difference, which is perceptible in the scattering phases, is the origin of the difference in strengths observed. Charge symmetry breaking is very significant in the lambda-nucleon force \cite{har1971, gal2015, gal2018}.

As a test to validate this application of GLM theory, phase shifts were computed from the new potentials and compared with the original phase shifts from the NSC1997f potential. Comparisons of these phase shifts are shown in Figures \ref{fig:compare_lp} and \ref{fig:compare_ln}. It can be seen that the potentials constructed through GLM theory are phase equivalent to the original NSC1997f potentials. Phase equivalence and the property of unitarity in $\tilde{S}_{0}(k)$, as verified in Section 7, are important checks on the accuracy of the application of GLM theory.  

\subsection{Effective potentials: spin-averaging across channels}

The total spin in a lambda-proton pair is either 0 or 1. When the total spin is 0, the spin multiplicity is 1 (singlet state), with a spin projection quantum number of 0. However, when the total spin is 1, the spin multiplicity is 3 (triplet of degenerate states) with projection quantum numbers -1, 0 and +1. Therefore, in a lambda-proton interaction the pair has a 1/4 probability of being in a singlet state and a 3/4 probability of being in a triplet state. These probabilities are are based on the assumption that each state is \textit{equally} likely. Thus, the \textit{effective} lambda-proton two-body potential, $V_{\Lambda p}$, is given by \cite{ram1964, ans1986}
\begin{align}
\label{eq:spinlp}
V_{\Lambda p} = \frac{1}{4}V_{\Lambda p}(^1\text{S}_0) + \frac{3}{4}V_{\Lambda p}(^3\text{S}_1)
\end{align} 

\begin{figure}[!h]
\centering
\includegraphics[width=0.9\linewidth]{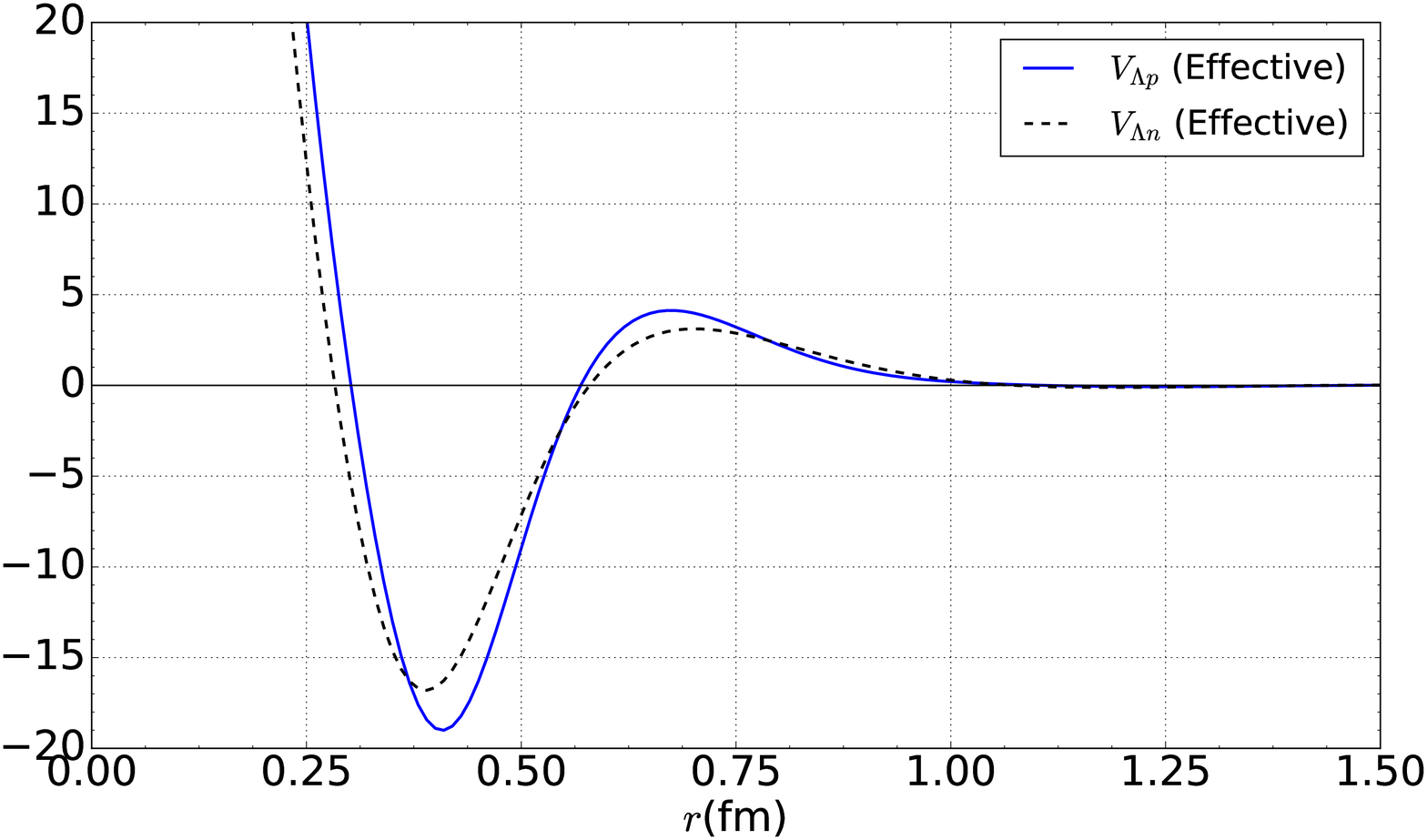}
\caption[]{Spin-averaged (effective) $\Lambda p$ and $\Lambda n$ two-body potentials. The spin average is computed as the sum of one-quarter of the singlet channel potential and three-quarters of the triplet channel potential. The attraction depth is -19.01 MeV for $\Lambda p$ and -16.80 MeV for  $\Lambda n$.}
\label{fig:both_effective} 
\end{figure}

Along similar lines of reasoning, the effective lambda-neutron two-body potential, $V_{\Lambda n}$, is obtained as follows:
\begin{align}
\label{eq:spinln}
V_{\Lambda n} = \frac{1}{4}V_{\Lambda n}(^1\text{S}_0) + \frac{3}{4}V_{\Lambda n}(^3\text{S}_1)
\end{align} 

This spin-averaging scheme was applied to the lambda-nucleon potentials constructed through GLM theory. The resulting effective two-body potentials are shown in Figure \ref{fig:both_effective}. The attraction depth in the effective lambda-proton potential is -19.01 MeV while that of the lambda-neutron potential is -16.80 MeV. As discussed earlier, this stronger attraction in the lambda-proton potential is the result of charge symmetry breaking in the lambda-nucleon force.

\section{Conclusions}

In this paper, new energy-independent lambda-proton and lambda-neutron potentials that are unique were recovered from theoretical scattering data below the inelastic threshold. These potentials were restored through Gel'fand-Levitan-Marchenko theory. The general features of a baryon-baryon interaction, short-range repulsion and intermediate-range attraction, can be observed in these potentials. Furthermore, charge symmetry breaking that is discernible in the lambda-nucleon scattering data is preserved in the new potentials: the lambda-proton force is slightly more attractive than the lambda-neutron force. This effortless inclusion of charge symmetry breaking in inverse scattering theory is an advantage, when compared with other theories where separate elaborate schemes must be implemented. In summary, these new potentials are energy-independent, unique and conform to charge symmetry breaking, in addition to being spin-dependent. Assessments on the accuracy of these potentials can be carried out through few-body calculations on lambda hypernuclei. It is important to emphasize that $\Lambda N - \Sigma N$ coupling was not incorporated into these potentials. At the moment, there are no known mechanisms for including $\Lambda N - \Sigma N$ coupling in inverse scattering theory. 

\section*{References}

\bibliography{nuclearref}{}

\providecommand{\newblock}{}
\begin{thebibliography}{10}
\expandafter\ifx\csname url\endcsname\relax
  \def\url#1{{\tt #1}}\fi
\expandafter\ifx\csname urlprefix\endcsname\relax\def\urlprefix{URL }\fi
\providecommand{\eprint}[2][]{\url{#2}}

\bibitem{dav2005}
Davis D~H 2005 {\em Nuclear Physics A\/} {\bf 754} 3 -- 13 proceedings of the
  Eighth International Conference on Hypernuclear and Strange Particle Physics

\bibitem{dal2005}
Dalitz R~H 2005 {\em Nuclear Physics A\/} {\bf 754} 14 -- 24 proceedings of the
  Eighth International Conference on Hypernuclear and Strange Particle Physics

\bibitem{deSwart1971}
de~Swart J~J, Nagels M~M, Rijken T~A and Verhoeven P~A 1971 {\em Springer
  Tracts in Modern Physics\/} vol \textbf{60} ed H{\"o}hler G (Berlin,
  Heidelberg: Springer Berlin Heidelberg) pp 138--203

\bibitem{deSwart1996}
de~Swart J~J, Klomp R~A~M~M, Rentmeester M~C~M and Rijken T~A 1996 {\em
  Few-Body Problems in Physics '95: In memoriam Professor Paul Urban\/} ed
  Guardiola R (Vienna: Springer Vienna) pp 438--447

\bibitem{rij1993}
Rijken T~A 1993 {\em Few-Body Problems in Physics '93, Supplementum 7.
  Proceedings of the XIVth European Conference on Few-Body Problems in
  Physics\/} ed Bakker B~L~G and van Dantzig R (Wien: Springer Verlag) pp 1--12

\bibitem{rij2001}
Rijken T~A 2001 {\em Nuclear Physics A\/} {\bf \textbf{691}} 322 -- 328 proc.
  7th Int. Conf. on Hypernuclear and Strange Particle Physics

\bibitem{hol1989}
Holzenkamp B, Holinde K and Speth J 1989 {\em Nuclear Physics A\/} {\bf
  \textbf{500}} 485 -- 528

\bibitem{reu1992}
Reuber A, Holinde K and Speth J 1992 {\em Czechoslovak Journal of Physics\/}
  {\bf \textbf{42}} 1115--1135

\bibitem{hai2005}
Haidenbauer J and Mei\ss{}ner U~G 2005 {\em Physical Review C\/} {\bf
  \textbf{72}} 044005

\bibitem{fuj1996a}
Fujiwara Y, Nakamoto C and Suzuki Y 1996 {\em Physical Review C\/} {\bf
  \textbf{54}}(5) 2180--2200

\bibitem{fuj1996b}
Fujiwara Y, Nakamoto C and Suzuki Y 1996 {\em Physical Review Letters\/} {\bf
  \textbf{76}}(13) 2242--2245

\bibitem{fuj2001}
Fujiwara Y, Kohno M, Nakamoto C and Suzuki Y 2001 {\em Phys. Rev. C\/} {\bf
  64}(5) 054001

\bibitem{pol2006}
Polinder H, Haidenbauer J and Mei{\ss}ner U~G 2006 {\em Nuclear Physics A\/}
  {\bf 779} 244 -- 266

\bibitem{pol2007}
Polinder H, Haidenbauer J and Mei{\ss}ner U~G 2007 {\em Physics Letters B\/}
  {\bf 653} 29 -- 37

\bibitem{hai2013}
Haidenbauer J, Petschauer S, Kaiser N, Mei{\ss}ner U~G, Nogga A and Weise W
  2013 {\em Nuclear Physics A\/} {\bf 915} 24 -- 58

\bibitem{gal2018b}
Gal A and Garcilazo H 2019 {\em Physics Letters B\/} {\bf 791} 48 -- 53

\bibitem{ada2018}
Adamczyk L {\em et~al.\/} 2018 {\em Phys. Rev. C\/} {\bf 97}(5) 054909 (STAR
  Collaboration)

\bibitem{mar1950}
Marchenko V~A 1950 {\em Dokl. Akad. Nauk SSSR\/} {\bf 72} 457--460

\bibitem{mar1955}
Marchenko V~A 1955 {\em Dokl. Akad. Nauk. SSSR\/} {\bf 104} 695 -- 698

\bibitem{agr1963}
Agranovich Z~S and Marchenko V~A 1963 {\em The Inverse Problem of the
  Scattering Theory\/} (New York: Gordon and Breach)

\bibitem{bor1946}
Borg G 1946 {\em Acta Math.\/} {\bf 78} 1--96 (In German)

\bibitem{bor1949}
Borg G 1949 {\em Acta Math.\/} {\bf 81} 265--283

\bibitem{pov1948}
Povzner A 1948 {\em Mat. Sb. (N.S.)\/} {\bf 23} 3 -- 52 (In Russian)

\bibitem{lev1949}
Levitan B~M 1949 {\em Uspekhi Mat. Nauk\/} {\bf 4} 3 -- 112 (In Russian)

\bibitem{bar1949}
Bargmann V 1949 {\em Rev. Mod. Phys.\/} {\bf 21}(3) 488--493

\bibitem{bar1949b}
Bargmann V 1949 {\em Phys. Rev.\/} {\bf 75}(2) 301--303

\bibitem{lev1949b}
Levinson N 1949 {\em Matematisk Tidsskrift. B\/}  25--30

\bibitem{lev1949c}
Levinson N 1949 {\em Danske Vid. Selsk. Mat.-Fys. Medd.\/} {\bf 25} 29

\bibitem{lev1987}
Levitan B~M 1987 {\em Inverse Sturm-Liouville Problems\/} (Utrecht: VNU Science
  Press) (Translated from Russian by O Efimov)

\bibitem{del1938a}
Delsarte J 1938 {\em C.R. Acad. Sci. Paris\/} {\bf \textbf{206}} 178 -- 182 (In
  French)

\bibitem{del1938b}
Delsarte J 1938 {\em Journal de Mathematiques Pures et Appliques\/} {\bf
  \textbf{17}} 213 -- 231 (In French)

\bibitem{mar1952}
Marchenko V~A 1952 {\em Trudy Moskov. Mat. Ob\v{s}\v{c}.\/} {\bf 1} 327--420
  ISSN 0134-8663

\bibitem{coz1973}
Coz M and Coudray C 1973 {\em Journal of Mathematical Physics\/} {\bf 14}
  1574--1578

\bibitem{coz1987}
Coz M, Kuberczyk J and von Geramb H~V 1987 {\em Zeitschrift f{\"u}r Physik A
  Atomic Nuclei\/} {\bf \textbf{326}} 345--351

\bibitem{mas1999}
Massen S~E, Sofianos S~A, Rakityansky S~A and Oryu S 1999 {\em Nuclear Physics
  A\/} {\bf 654} 597 -- 611

\bibitem{ben1969}
Benn J and Scharf G 1969 {\em Nuclear Physics A\/} {\bf 134} 481 -- 504

\bibitem{ben1972}
Benn J and Scharf G 1972 {\em Nuclear Physics A\/} {\bf \textbf{183}} 319 --
  336

\bibitem{new2002}
Newton R~G 2002 {\em Scattering theory of waves and particles\/} 2nd ed (New
  York: Dover)

\bibitem{cha1977}
Chadan K and Sabatier P~C 1977 {\em Inverse problems in quantum scattering
  theory\/} (New York: Springer-Verlag)

\bibitem{sof1990}
Sofianos S~A, Papastylianos A, Fiedeldey H and Alt E~O 1990 {\em Physical
  Review C\/} {\bf \textbf{42}}(2) R506--R509

\bibitem{har1981}
Hartt K 1981 {\em Phys. Rev. C\/} {\bf 23}(6) 2399--2404

\bibitem{rak2007}
Rakityansky S~A, Sofianos S~A and Elander N 2007 {\em Journal of Physics A:
  Mathematical and Theoretical\/} {\bf 40} 14857

\bibitem{pap1990}
Papastylianos A, Sofianos S~A, Fiedeldey H and Alt E~O 1990 {\em Phys. Rev.
  C\/} {\bf 42}(1) 142--148

\bibitem{kub1987}
Kuberczyk J, Coz M, von Geramb H~V and Lumpe J~D 1987 {\em Zeitschrift f{\"u}r
  Physik A Atomic Nuclei\/} {\bf 328} 265--273

\bibitem{kir1989}
Kirst T, Amos K, Berge L, Coz M and von Geramb H~V 1989 {\em Physical Review
  C\/} {\bf \textbf{40}}(2) 912--923

\bibitem{fad1959}
Faddeev L~D 1959 {\em Uspekhi Mat. Nauk\/} {\bf 14}(4) 57--119

\bibitem{fad1963}
Faddeyev L~D and Seckler B 1963 {\em Journal of Mathematical Physics\/} {\bf 4}
  72--104

\bibitem{oli2014}
Olive K~A {\em et~al.\/} 2014 {\em Chinese Physics C\/} {\bf 38} 090001
  (Particle Data Group)

\bibitem{ren2017}
Rentmeester M~C~M and Klomp R~A~M {NN} {O}nline http://nn-online.org/ accessed:
  2017 May 01

\bibitem{rij1999}
Rijken T~A, Stoks V~G~J and Yamamoto Y 1999 {\em Physical Review C\/} {\bf
  \textbf{59}}(1) 21--40

\bibitem{hau1977}
Hauptman J~M, Kadyk J~A and Trilling G~H 1977 {\em Nuclear Physics B\/} {\bf
  125} 29 -- 51

\bibitem{alt1994}
Alt E~O, Howell L~L, Rauh M and Sofianos S~A 1994 {\em Phys. Rev. C\/} {\bf
  49}(1) 176--187

\bibitem{amo1989}
Amos K, Berge L, Geramb H~V~V and Coz M 1989 {\em Nuclear Physics A\/} {\bf
  499} 45 -- 62

\bibitem{hiy2002}
Hiyama E, Kamimura M, Motoba T, Yamada T and Yamamoto Y 2002 {\em Physical
  Review C\/} {\bf \textbf{66}}(2) 024007

\bibitem{how1993}
Howell L~L, Sofianos S~A, Fiedeldey H and Pantis G 1993 {\em Nuclear Physics
  A\/} {\bf \textbf{556}} 29 -- 41

\bibitem{lee1992}
Leeb H and Leidinger D 1992 {\em Few-Body Problems in Physics\/} ed Ciofi~degli
  Atti C, Pace E, Salm{\`e} G and Simula S (Vienna: Springer Vienna) pp
  117--127 ISBN 978-3-7091-7581-1

\bibitem{adam1993b}
Adam R~M, Fiedeldey H, Sofianos S~A and Leeb H 1993 {\em Nuclear Physics A\/}
  {\bf 559} 157 -- 172

\bibitem{nag2019}
Nagels M~M, Rijken T~A and Yamamoto Y 2019 {\em Phys. Rev. C\/} {\bf 99}(4)
  044002 \urlprefix\url{https://link.aps.org/doi/10.1103/PhysRevC.99.044002}

\bibitem{meo2017}
Meoto E~F and Lekala M~L 2017 {\em Journal of Physics: Conference Series\/}
  {\bf 915} 012004

\bibitem{har1971}
Hartt K and Sullivan E 1971 {\em Phys. Rev. D\/} {\bf 4}(5) 1353--1366

\bibitem{gal2015}
Gal A 2015 {\em Physics Letters B\/} {\bf 744} 352 -- 357

\bibitem{gal2018}
Gal A and Gazda D 2018 {\em Journal of Physics: Conference Series\/} {\bf 966}
  012006

\bibitem{ram1964}
Ram B and Downs B~W 1964 {\em Phys. Rev.\/} {\bf 133}(2B) B420--B428

\bibitem{ans1986}
Ansari H~H, Shoeb M and Khan M~Z~R 1986 {\em Journal of Physics G: Nuclear
  Physics\/} {\bf 12} 1369

\end{thebibliography}
\bibliographystyle{iopart-num}

\end{document}